\documentclass[twocolumn,showpacs,preprintnumbers,amsmath,amssymb,groupedaddress,prl]{revtex4}

\usepackage{graphicx}
\usepackage{dcolumn}
\usepackage{bm}
\usepackage{epsfig}
\usepackage{subfigure}
\def\e{\begin{equation}}
\def\f{\end{equation}}
\def\=#1{\overline{\overline #1}}

\def\_#1{{\bf #1}}
\def\.{\cdot}
\def\l#1{\label{eq:#1}}
\def\r#1{(\ref{eq:#1})}

\def\o{\omega}
\def\O{\Omega}
\def\va{\varepsilon}
\def\o{\omega}
\def\O{\Omega}
\def\.{\cdot}
\def\x{\times}

\begin{document}

\title{Bianisotropic route to the realization and matching of backward-wave
metamaterial slabs}

\author{S. A. Tretyakov,$^1$ C. R. Simovski,$^{1,2}$ M. Hudli\v{c}ka$^{1,3}$}
\affiliation{$^1$Radio Laboratory/SMARAD, Helsinki University of
Technology, P.O. Box 3000, FI-02015 TKK, Finland\\
$^2$Physics Department, State University of Information
Technologies, Mechanics and Optics, Sablinskaya 14, 197101, St.~Petersburg, Russia\\
$^3$Faculty of Electrical Engineering, Czech Technical University,
Technick\'{a} 2, 16627, Prague 6, Czech Republic}

\date{\today}
\begin{abstract}
A concept of backward-wave bianisotropic composite medium matched to
free space is suggested. It is based on the use of a uniaxial
bianisotropic structure embedded into a matrix with negative
effective permittivity. Since bianisotropy is easier to achieve in
the optical range than artificial magnetism, this concept is
prospective for optical backward-wave metamaterials. As an example
of possible realizations, a microwave $\Omega$-composite combined
with a wire lattice is analytically studied.
\end{abstract}

\pacs{78.20.Ci, 42.70.Qs, 42.25.Gy, 73.20.Mf, 78.67.Bf}

\maketitle




Design and studies of materials with negative electromagnetic
parameters supporting backward waves is currently a very active
field of research. The concept of backward waves is not new: It goes
back to the beginning of the 19th century and is connected to the
names of Lamb, Schuster, and Pocklington. In the middle of the 20th
century, this concept was extended to waves in homogeneous materials
and negative refraction effect was theoretically predicted
(Mandelshtam, Sivukhin, Silin, Veselago). Detailed reviews on the
current status and on the history of this research field can be
found in e.g.
\cite{Shalaev_r,Lapin,Veselago_Narimanov,rev1,rev3,4,rev2}.

 The main application for backward-wave materials is in
sub-wavelength imaging devices (``perfect lens'' \cite{1}-\cite{3}),
but a full range of other possibilities is expected, especially for
optical frequencies, if low-loss and matched slabs of backward-wave
materials are realized \cite{4,5}. The well-known design approach
for the microwave range is based on the use of arrays of long thin
metal conductors and split rings \cite{6}. This approach has been
extended to terahertz \cite{7} and even infrared frequency range
\cite{11}. Realization of negative permeability with the use of
split rings becomes very problematic at optical frequencies, and
some alternative approaches have been proposed \cite{8}-\cite{10}.
Considerable progress has been reported along this route, but the
samples realized so far suffer from high losses and poor matching
with free space \cite{7,11}. There exist possibilities to realize
backward waves also in anisotropic media \cite{Ismo,Pavel} and
in anisotropic waveguides \cite{Podol}, which do not necessarily
require magnetic properties of materials, but are restricted to
strongly anisotropic structures. In addition, difficulties to
realize backward-wave samples {\itshape matched} to free space
inhibit potential applications both in the microwave and in the
optical regions.

Majority of researchers focus on the design of
magneto-dielectrics, where  the backward-wave regime is realized
when both the permittivity  $\epsilon$ and permeability $\mu$ have
negative real parts. Following to paper \cite{14}, the
bianisotropy is usually considered as a factor that one should
avoid in the design of backward-wave materials, and effort is
often concentrated on the design of symmetrical variants of split
rings which minimizes magnetoelectric coupling \cite{Souk}.
However, backward waves can exist in more general linear media,
namely in bianisotropic media (e.g., in chiral media
\cite{nih}-\cite{13}). It was recently demonstrated that in
chiral media it is possible to improve the characteristics of
backward-wave regime \cite{13,He}. Moreover, as we will show in
this paper, there is a possibility to benefit from more design
freedom offered by additional material parameters. Here we will
show that it is possible to design a bianisotropic material in
such a way that it supports linearly-polarized backward waves and
a slab made of this material is perfectly matched to free space
for the normal direction of propagation.

The material can be realized as a composite with $\Omega$-shaped
metal inclusions, and can be called \emph{omega-medium}
\cite{Engheta}. Since  reliable analytical models of
$\Omega$-particles and $\Omega$-media were developed and checked
numerically and experimentally for the microwave range
\cite{JEWA,EM,Serdyuk}, we use these models (valid below
$70-100$ GHz) for a demonstration of the concept having in mind an
optical realization as one of targets.

Reciprocal uniaxial $\Omega$-media obey the following constitutive
relations \cite{Serdyuk}: \e
\_D=\=\epsilon\.\_E+j\sqrt{\varepsilon_0\mu_0} K\=J\.\_H ,\quad
\_B=\=\mu\.\_H+j\sqrt{\varepsilon_0\mu_0} K\=J\.\_E\f Denoting the
unit vector along the optical axis as $\_z_0$ we can write the
permittivity and permeability dyadics in the form: \e
\=\epsilon=\va_0\left(\varepsilon_t\=I_t+\varepsilon_z
\_z_0\_z_0\right),\quad \=\mu=\mu_0\left(\mu_t\=I_t+\mu_z
\_z_0\_z_0\right),\f where $\=I_t$ is the two-dimension unit
dyadic defined in the plane orthogonal to $\_z_0$:
$\=I_t=\_x_0\_x_0+\_y_0\_y_0$. The magnetoelectric dyadic is
antisymmetric: $ \=J=\_z_0\x\=I_t$.  Complex dimensionless
parameter $K$ measures the magnetoelectric coupling effect.
Eigenwaves in such media are linearly polarized plane waves,
similarly to simple magneto-dielectrics.

In this paper we consider a slab of such composite material
with the optical axis orthogonal to the slab surface. For simplicity
of the analysis we concentrate on the normal-incidence excitation.
In this case both eigenwaves in the slab have the same propagation
constant \cite{Serdyuk}
\e
\beta =k_0 \sqrt {{\varepsilon _t \mu _t } - K^2 }
,\l{beta}\f where $k_0$ is the free-space wavenumber.
For the characteristic admittances which are different for the
waves traveling in the positive and negative directions of the
axis $z$ (denoted below as $Y_+$ and $Y_-$, respectively), the
following relation holds \cite{Serdyuk}:
\e Y_ \pm = Y_0 \sqrt
{\frac{{\varepsilon _t }}{{\mu _t }}} \left( {\sqrt {1 - {K^2\over
\va_t\mu_t} } \mp j{K\over \sqrt{\va_t\mu_t}}} \right), \l{yyy}\f
where $Y_0 = \sqrt {\varepsilon _0 /\mu _0 }$ is the free-space
wave admittance.

In \cite{Sergei1} it was shown that omega composites can be used to
realize absorbing layers which are matched to free space. Indeed,
one can notice that if the material parameters satisfy the condition
\e K= \frac{{ j }}{2}\left( {\mu _t  - \varepsilon _t } \right),
\l{cond1} \f relation \r{yyy} gives $Y_+=Y_0$. Then the reflection
coefficient from one side of the slab of an arbitrary thickness $d$ equals $R=0$
and the transmission coefficient reads $ T = \exp(- j\beta d)$. Note
that the matching condition can be satisfied for arbitrary
permittivity and permeability values, provided one can control the
coupling coefficient $K$.

If condition \r{cond1} is satisfied, the propagation constant of the
eigenwaves in the medium reads \r{beta} \e \beta= \frac{{k_0
}}{2}\left( {\varepsilon _t + \mu _t } \right). \l{kx0} \f This
result shows that the magnetoelectric coupling not only allows one
to match the slab to free space, but also makes it easier to realize
the backward-wave regime. From formula \r{kx0} we see that the
refractive index in a matched $\Omega$-slab is negative when \e {\rm
Re}\{\varepsilon_t+\mu_t\}<0,\l{cond2}\f which is easier to satisfy
than the usual conditions \e {\rm Re}\{\varepsilon_t\}<0,\qquad {\rm
Re}\{\mu_t\}<0.\f In particular, it is not necessary to realize
negative permeability, which can be more difficult at the optical
frequencies than to satisfy \r{cond1}. Composites with negative
permittivity and acceptable losses, on the contrary, are available
in optics and can be obtained, for example, using dilute arrays of
metal nanoparticles embedded in a dielectric matrix.
Like metal split rings \cite{7,11,Souk}, metal bianisotropic
particles can possess a resonance in the infrared range and,
possibly, in the visible. Resonant magnetoelectric coupling of these
particles can be sufficient even if the corresponding resonant magnetic
polarizability is not high. This expectation is based on the fact that the effect
of artificial permeability  is of the
order of $\mu\sim O(k_0d)^2$, while the effect of magneto-electric
coupling is an order of magnitude stronger: $K\sim O(k_0d)$. Here
$d$ is the characteristic particle size.

\begin{figure}[h!]
\centering \epsfig{file=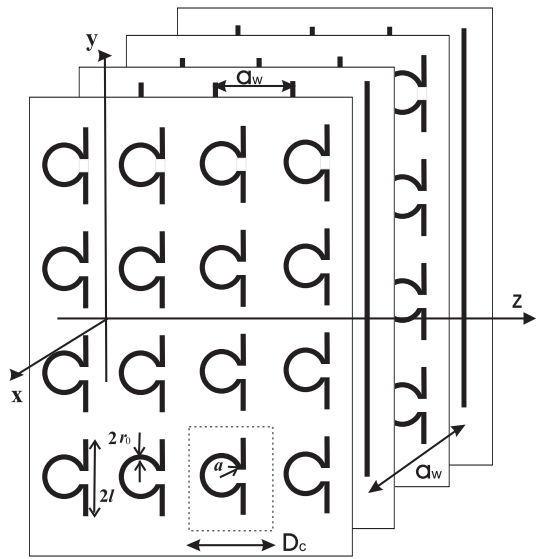, width=7cm}
\caption{Geometry of the matched backward-wave structure and of an
individual particle. The uniaxial structure has a second
identical set of particles lying in the $x-z$ plane (not shown),
however that second set does not interact with the eigenwave with
the electric field polarized along $y$. $D_c$ is the $\O-$lattice
unit cell size, $a_w$ is the period of the wire lattice.}
\label{figa}
\end{figure}

To study if a realization is possible with a particular realistic dimensions we
will next consider an example for the microwave frequency region.
Negative permittivity at microwaves can be obtained using a lattice
of parallel wires (e.g., \cite{Pendry}). A uniaxial omega composite
can be realized as a lattice of pairs of $\O$-particles. One
particle in the pair is orthogonal to the other
\cite{Sergei1,Serdyuk}. In this paper we restrict our study by the
case of the normal propagation direction. Then, for the simplified structure
shown in Fig.~\ref{figa} all the formulas written above for a uniaxial
composite are valid.
Our purpose is to show that conditions \r{cond1} and \r{cond2}
correspond to physically achievable geometric parameters of
$\O$-particles and wires forming the composite medium shown in Fig.~\ref{figa}.

The microwave electromagnetic model of an omega particle represents
the particle as a connection of a short wire dipole antenna and a
small loop antenna \cite{JEWA}.  Small bianisotropic particles can
be characterized by dyadic electric and magnetic polarizabilities,
which define the relations between induced electric and magnetic
dipole moments \textbf{p}, \textbf{m} and external electric and
magnetic fields \textbf{E}, \textbf{H} applied to the particle
\cite{Engheta}--\cite{Serdyuk}: \e \textbf{p} = \overline {\overline
\alpha } _{ee} \.\textbf{E} + \overline {\overline \alpha  } _{em}
\.\textbf{H} \f \e \textbf{m} = \overline {\overline \alpha  } _{me}
\.\textbf{E} + \overline {\overline \alpha } _{mm} \.\textbf{H} \f

For electrically small particles a lumped-element equivalent circuit
models of such antennas give acceptable accuracy \cite{Serdyuk}, and
the particle polarizabilities can be expressed in terms of
equivalent circuit parameters. If the loop radius $a$ of an
$\Omega$-particle shown in Fig.~\ref{figa} is equal to or larger
than the arm length $l$, the polarizability components relevant in
the case of the normal propagation can be approximated as \e \alpha
_{ee}^{yy} = \frac{A}{{\omega _0^2 - \omega ^2 + j \omega \Gamma}},
~~~~~~A = {l^2\over L_0} \l{C1aeezz},\f \e \alpha _{mm}^{xx} =
\frac{{B\omega ^2 }}{{\omega _0^2  - \omega ^2 + j \omega \Gamma}},
~~~~~~B = \frac{{\pi^2\mu_0 ^2 a^4 }}{{L_0 }}, \l{C1ammzz} \f \e
\alpha _{me}^{xy} = \frac{{ { j }\omega D}}{{\omega _0^2 - \omega ^2
+ j \omega \Gamma}}, ~~~~~~D =\frac{{2\pi^2\mu_0\va_0\va_m
a^4}}{{L_0C_0 }}, \l{C1amezz} \f
where $\omega _0  = 1/\sqrt {L_0
C_0 }$, $\Gamma = R_l /L_0$, and $\varepsilon_m$ is the matrix relative permittivity.
The matrix is assumed to be non-magnetic: The matrix permeability
$\mu_r=1$.
The equivalent circuit parameters
read \e R_l = \sqrt {\frac{{\omega \mu_0 }}{{2\sigma }}}
\frac{a}{{r_0 }},\qquad L_0 = \mu_0 a\left[ {\log \left(
{\frac{{8a}}{{r_0 }}} \right) - 2} \right],\l{param1} \f \e C_0 =
\frac{{\pi l\varepsilon_0 \varepsilon_m }}{{\log \left( {2l/r_0 }
\right)}}, \l{param}\f where $\sigma$ is the metal conductivity,
and dimensions
$r_0$, $a$ and $l$ are defined in Fig.~\ref{figa}. Here we have
neglected the electric polarization of the ring compared to that of
the dipole portion which is an acceptable approximation for
electrically small $\O$-particles \cite{Serdyuk}. However, the
bianisotropy of the particle described by formula \r{C1amezz} is
practically that of the split ring loaded by a lumped capacitance
$C_0$. This approximation is suitable for the case $a\ge l$.

For the relative transversal permittivity and permeability and for
the magnetoelectric coupling the low-density approximation (the
particles concentration $N$ is such that the unit cell of the
composite is significantly larger than the particles sizes) gives in
accordance with \cite{Sergei1}:

\e \varepsilon _t  = \varepsilon _r + \frac{{N\alpha _{ee}^{yy}
}}{{\varepsilon _0 }} = \varepsilon _r + \frac{{NA}}{{\varepsilon _0
\left( {\omega _0^2  - \omega ^2 + j \omega \Gamma} \right)}},
\l{epseff} \f \e \mu _t  = \mu _r  + \frac{{N\alpha _{mm}^{xx}
}}{{\mu _0 }} = \mu _r  + \frac{{NB\omega ^2 }}{{\mu _0 \left(
{\omega _0^2 - \omega ^2 + j \omega \Gamma} \right)}}, \l{mueff} \f
\e K =   N j \frac{{\alpha _{me}^{xy} }}{{\sqrt {\varepsilon _0 \mu
_0 } }} =- N\frac{{\omega D}}{{\left( {\omega _0^2  - \omega ^2 + j
\omega \Gamma} \right)\sqrt {\varepsilon _0 \mu _0 } }}, \l{K1} \f
where $\varepsilon_r$ and $\mu_r$ are the relative effective
permittivity and permeability of the background material,
respectively. In formulas \r{epseff}-\r{K1} we use a different
notation for the background material permittivity than in \r{param}
because in our structure $\va_r\ne \va_m$. The background
permittivity is, of course, affected by the presence of metal wires
shown in Fig.~\ref{figa}. However, the capacitance $C_0$ is
determined by the quasi-stationary electric field in the small
spatial domain of the particle.  That is why the matrix permittivity
$\va_m$ and not $\va_r$ enters $C_0$ (see more on this important
issue in \cite{paradox} and \cite{Kluwer}).

Conditions \r{cond1} and \r{cond2} should be satisfied at the same
operational frequency $\o$. Substituting \r{epseff}-\r{K1} into
\r{cond1} we come to equations for the real and imaginary parts of
\r{cond1}: \e {2D\over \sqrt{\va_0\mu_0}}= {(1-\va_r)\over
N}\Gamma,\l{first}\f \e {(1-\va_r)\over
N}(\o_0^2-\o^2)=\left({A\over \va_0}-{B\o^2\over
\mu_0}\right)\l{second}.\f

Substituting \r{epseff}-\r{K1} into \r{cond2} together with
\r{first} and \r{second} it is possible to show that
condition $\varepsilon_r<-1$ should hold
for simultaneous validity of  conditions \r{cond1}
and \r{cond2}. This inequality can be satisfied with
the negative background permittivity created by the wire medium. If the
wave propagates in the plane orthogonal to the wires and the
electric field is oriented along them, the wire array behaves as a
low-frequency plasma. The permittivity of the wire lattice is then
not spatially dispersive, and we have \cite{Pendry,Best} \e
\varepsilon_r ( \omega) = \va_m\left(1 - \frac{{\omega _p^2
}}{{\omega ^2 }}\right). \l{omp}\f In \cite{Best} one can find two
formulas for the plasma frequency of such wire arrays, the first
one, \e \omega _p^2 = \frac{{2\pi }}{{\va_0\mu_0 a_w^2 \left( \log
{a_w \over 2\pi r_w }+0.5275 \right)}}, \l{thin}\f is valid if the
radius $r_w$ of wires is very small (practically $r_w<a_w/100$).
If it is not so small ($a/20<r_w<a_w/5$), the second formula is
applicable: \e \omega _p^2 = \frac{{2\pi }}{{a_w^2 \va_0\mu_0\log
{a_w^2 \over 4 r_w(a_w - r_w) }}}.\l{thick}\f
\begin{figure}
 \centering
\subfigure[][]{\label{a}\includegraphics[width=4.2cm]{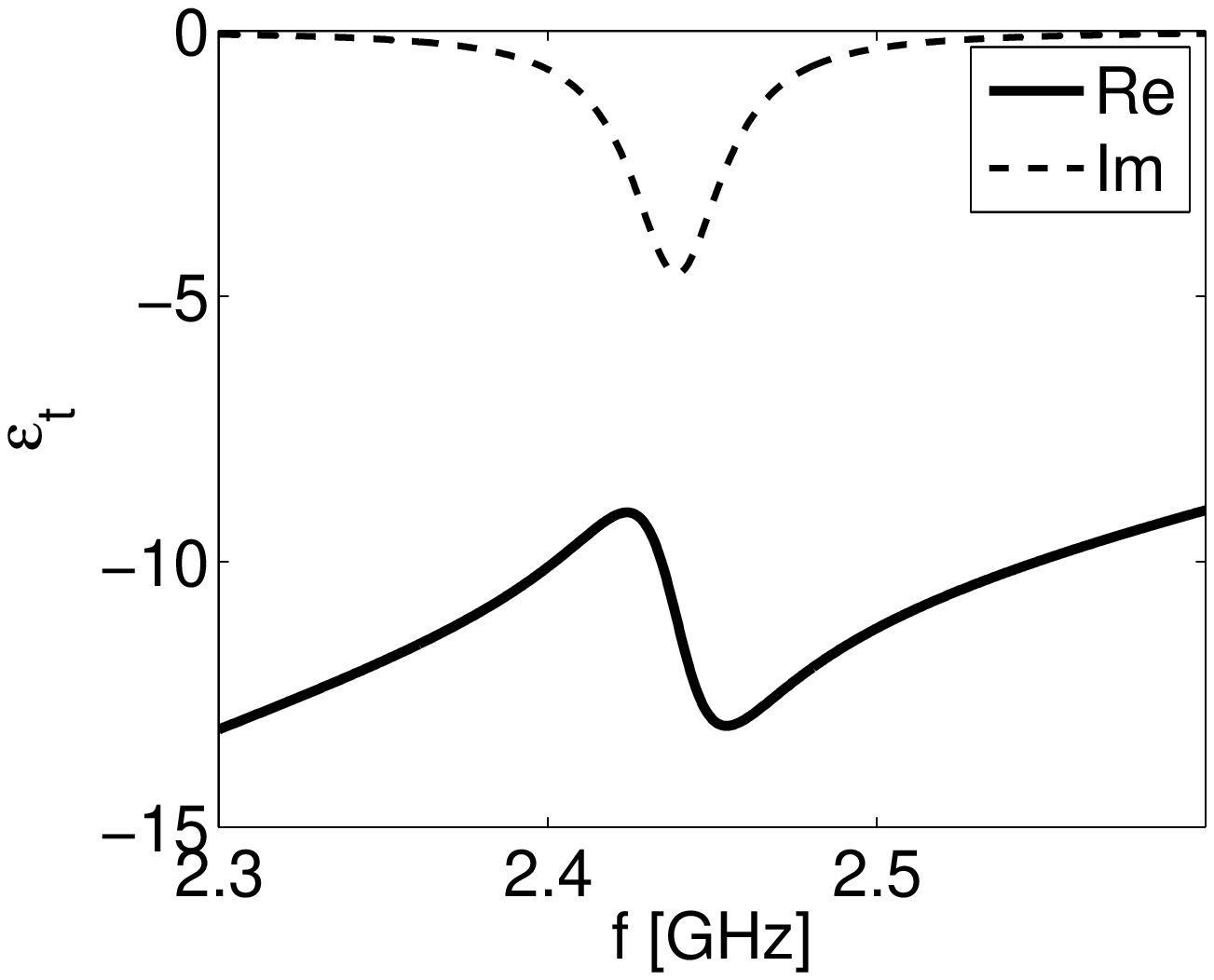}}
\subfigure[][]{\label{b}\includegraphics[width=4.2cm]{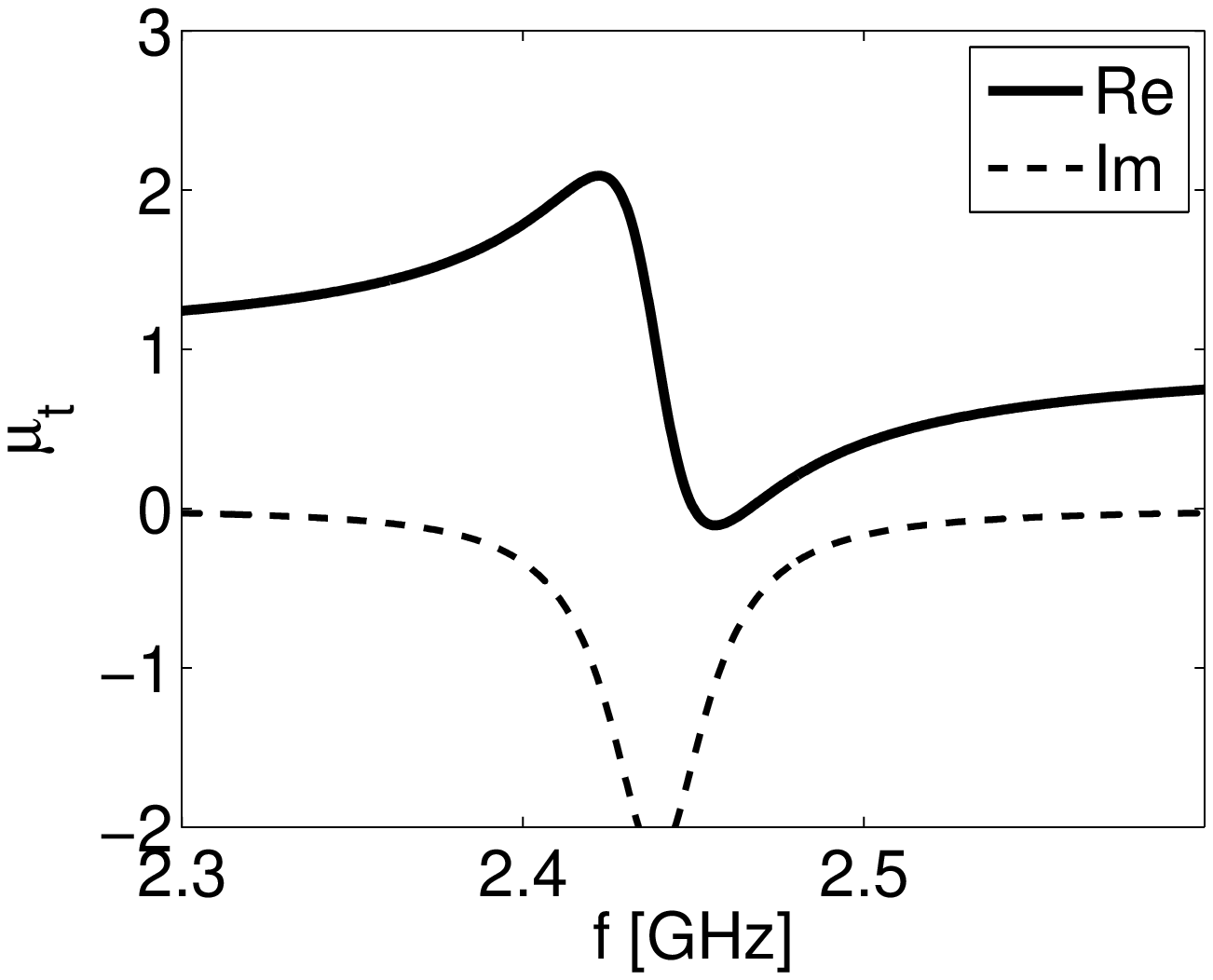}}
\subfigure[][]{\label{c}\includegraphics[width=4.25cm]{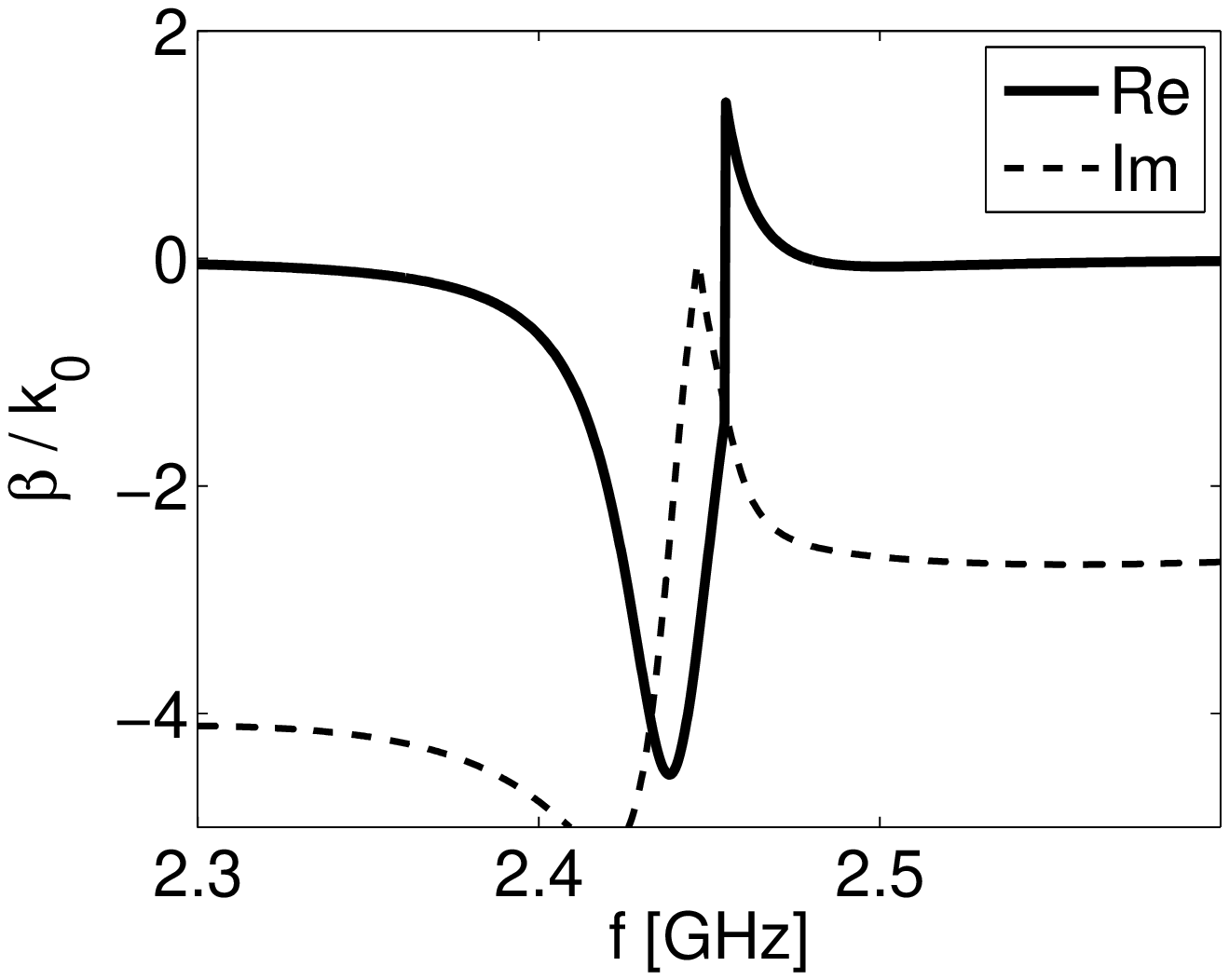}}
\subfigure[][]{\label{d}\includegraphics[width=4.25cm]{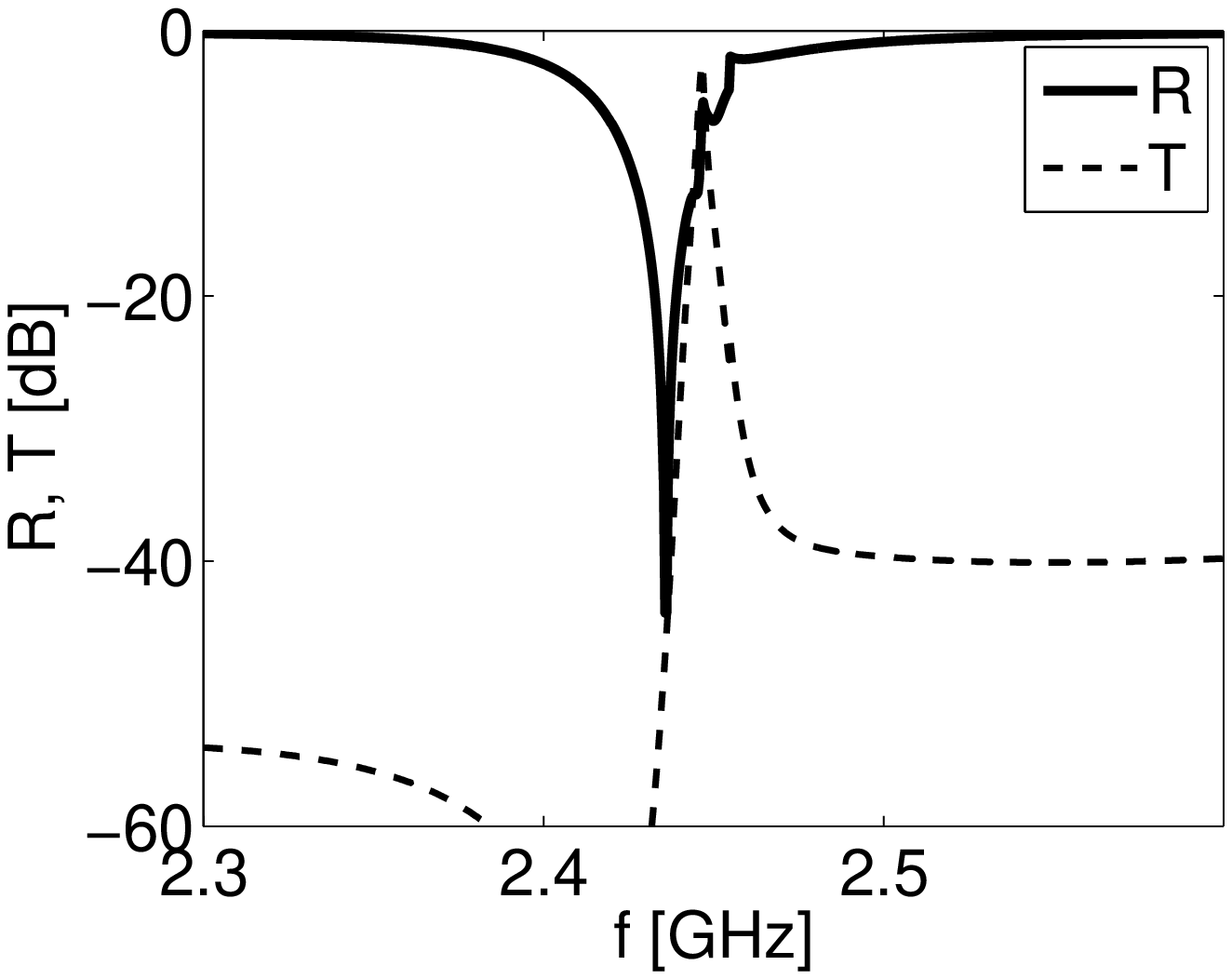}}
\caption{Parameters of the bianisotropic layer with the designed
lattice geometry. Effective permittivity \subref{a}, permeability
\subref{b}, normalized propagation factor $\beta/k_0$ \subref{c},
reflection and transmission coefficients in dB for a $32$-mm thick
layer \subref{d}.} \label{figa1}
\end{figure}

Let us pick up some reasonable values of parameters $\Gamma, \va_m,\
a,\ l,\ r_0$ (so that $a>l$ and $a,l\gg r_0$) and calculate $\o_0$.
Then from \r{first} and \r{second} we find $(1-\va_r)/N$ and $\o$.
At the frequency $\o$ the medium under design is perfectly matched.
Then varying $N$ we satisfy the condition $\va_r<-1$. Finally we
express the needed metal conductivity through $\Gamma$: \e \sigma =
\frac{{\omega \mu _0 }}{2}\left( {\frac{a}{{\Gamma L_0 r_0 }}}
\right)^2. \f The results for the particle parameters should satisfy
to the following physical conditions:
\begin{itemize}
\item Effective medium condition: $D_c<\lambda_{\rm eff}/2$, where
$\lambda_{\rm eff}=2\pi/{\rm Re}(\beta)$ is the effective
wavelength in the medium at the working frequency and
$D_c=N^{-1/3}$ is the size of a unit cell of the $\O$-lattice.
\item Low concentration condition: $D_p={\rm max}(2a,2l)<{\rm
min}(D_c,a_w)$. This means that particles do not touch each other
and the wires. \item Metal conductivity: $\sigma$ should be that
of a known metal \emph{or} alloy.
\end{itemize}
If any of these conditions are violated, one can vary the initial
parameters $\Gamma,\ \va_m,\ a,\ l,\ r_0$ and finally find their
proper combination. Then we put the wire medium period be equal to
the unit cell size $a_w=D_c$ and find the wire medium plasma
frequency and the wire radius $r_w$ from formula \r{omp} together
with \r{thin} or \r{thick}. These formulas allow us to study the
frequency properties of the designed structure.

A result obtained for the following design parameters: $\va_m=10$,
$l=3.5$ mm, $a=3.9$ mm, $r_0=0.05$ mm, is presented in
Fig.~\ref{figa1} for the frequency band centered at the particle
resonant frequency $f_0=2.4393$ GHz. The perfect matching holds at
the frequency $f=2.4359$ GHz. The parameter $\Gamma=0.02\cdot
\omega_0$ corresponds to the conductivity $\sigma\approx
1.6\cdot10^{6}\ {1/{\rm Ohm\cdot m}}$. With small metal losses
(those of copper) we could not satisfy the effective medium
condition. Other parameters were found as explained above:
$N=2.34\cdot 10^5\ {1/{\rm m^3}}$ that corresponds to $D_c=a_w\approx
16.23$ mm and $r_w=0.07$ mm that allows the backward-wave condition
over the particle resonance band. One can see that the permeability
is positive in the backward-wave region. The backward-wave regime is
due to the bianisotropy. The layer thickness was chosen $d=2D_c$
(the number of particles across the layer is enough to use the
effective medium approximation). The thickness resonances happen at
much higher frequencies than $f$ (in the region where $\va_r>0$).


To conclude, a bianisotropic composite supporting backward waves
can be
obtained with physically realizable parameters of inclusions.
This makes the bianisotropic route to matching a backward-wave
composite be prospective for microwave and hopefully for optical
metamaterials.
Although this matching is perfect for the
normal incidence only, the mismatch for oblique incidence
should be of the same order as for conventional isotropic slabs with $\mu/\epsilon=\mu_0/\epsilon_0$.
It is also important to notice that the omega slab matching is totally thickness independent
and holds even for a semi-infinite lattice.

\bigskip This work has been supported in part by the Grant Agency of the
Czech Republic under project 102/06/1106 "Metamaterials,
nanostructures and their applications" and by the Finnish Academy
via its Center-of-Excellence program.

\end{document}